\documentclass[twoside,reqno]{mbook}
\usepackage{epsfig}
\usepackage{amssymb,amsmath}
\usepackage{times}
\newcommand{\SU}[2]{\ensuremath{\mathrm{SU}^{ #1 }( #2 )}}
\newcommand{\Un}[2]{\ensuremath{\mathrm{U}^{ #1 }( #2 )}}

\newcommand{\Spn}[1]{\ensuremath{\mathrm{Sp}( #1 )}}

\newcommand{\spn}[1]{\ensuremath{\mathfrak{sp}( #1 )}}

\newcommand{\etal}{\emph{et al.}}

\newcommand{\half}{\ensuremath{\textstyle{\frac{1}{2}}}}

\newcommand{\IASs}{isobaric analog $0^+$ states}
\newcommand{\CDB}{CD-Bonn}
\newcommand{\CDBt}{CD-Bonn+3terms}
\newcommand{\Gm}{GXPF1}
\newcommand{\HQ}{\ensuremath{H_Q^\perp(2)}}	

\newcommand{\Hsp}{\ensuremath{H_{\spn{4}}}}

\newcommand{\HM}{\ensuremath{H_{M}}}

\newcommand{\fpg}{\ensuremath{1f_{5/2}2p_{1/2}2p_{3/2}1g_{9/2}} }

\newcommand{\flevel}{\ensuremath{1f_{7/2}} }
\newcommand{\fFive}{\ensuremath{1f_{5/2}} }
\newcommand{\plevels}{\ensuremath{2p_{1/2}2p_{3/2}} }

\newcommand{\upfp}{upper {\it fp}}

\begin{document}

\sloppy \raggedbottom

 \setcounter{page}{1}

\newpage
\setcounter{figure}{0}
\setcounter{equation}{0}
\setcounter{footnote}{0}
\setcounter{table}{0}
\setcounter{section}{0}



\title{Dynamical Symmetries Reflected in Realistic Interactions}

\runningheads{Dynamical Symmetries Reflected in Realistic 
Interactions}{K.D.~Sviratcheva, J.P.~Draayer, and J.P.~Vary}

\begin{start}
\author{K. D. Sviratcheva}{1},
\coauthor{J. P. Draayer}{1},
\coauthor{J. P. Vary}{2}

\address{Department of Physics and Astronomy, Louisiana State
University, Baton
Rouge, LA 70803, USA}{1}

\address{Department of Physics and Astronomy, Iowa State
University, Ames, IA 50011,
USA,\\
Lawrence Livermore National Laboratory, L-414, 7000 East
Avenue, Livermore,
California, 94551, USA, and\\
Stanford Linear Accelerator Center, MS81, 2575 Sand Hill
Road, Menlo Park,
California, 94025, USA}{2}

\begin{Abstract}
Realistic nucleon-nucleon (NN) interactions, derived within the framework
of meson theory or more recently in terms of chiral effective field theory,
yield new possibilities for achieving a unified microscopic description of
atomic nuclei. Based on spectral distribution methods, a comparison of these
interactions to a most general $\mathrm{Sp(4)}$ dynamically symmetric
interaction, which previously we found to reproduce well that part of the
interaction that is responsible for shaping pairing-governed isobaric analog
$0^+$ states, can determine the extent to which this significantly simpler
model Hamiltonian can be used to obtain an approximate, yet very good
description of low-lying nuclear structure. And furthermore, one can apply
this model in situations that would otherwise be prohibitive because of the
size of the model space. In addition, we introduce a $\mathrm{Sp(4)}$
symmetry breaking term by including the quadrupole-quadrupole interaction in
the analysis and examining the capacity of this extended model interaction
to imitate realistic interactions. This provides a further step towards
gaining a better understanding of the underlying foundation of realistic
interactions and their ability to reproduce striking features of nuclei such
as strong pairing correlations or collective rotational motion.
\end{Abstract}
\end{start}


\section[]{Introduction}

Spectral distribution theory \cite{FrenchR71,ChangFT71} is an excellent
framework for comparing the overall behavior of microscopic interactions and
uncovering fundamental  properties of realistic $NN$ potentials as well
as derivative effective interactions
\cite{ChangFT71,DraayerOP75,Potbhare77}. Likewise, this information can be
propagated beyond the defining two-nucleon system to nuclei with larger
numbers of nucleons \cite{ChangFT71} and  for higher values of isospin
\cite{HechtDraayer74}.  We search for the level of respect for selected
underlying symmetries
\cite{SDV06,SDV06b} such as the \Spn{4} symmetry 
\cite{SGD03stg,SGD04} of isovector
(like-particle and proton-neutron, $pn$) pairing correlations plus an 
isoscalar $pn$
force and the \SU{}{3} symmetry
\cite{Elliott} of collective rotational modes.  Such symmetry-respecting
microscopic model interactions can be used to probe the pairing and rotational
characteristics of a realistic interaction
\cite{Draayer73,HalemaneKD78,KotaPP80,CounteeDHK81}, which will reflect the
characteristic properties of the pairing (quadrupole) model Hamiltonian if both
interactions strongly correlate.
As these symmetries are clearly important
for certain spectral features (for example, observed pairing gaps and 
enhanced electric
quadrupole transitions), we have a tool for rapidly assessing the likely
success of these interactions for reproducing those spectral features.

Recent applications of the theory of spectral distributions also 
include quantum chaos,
nuclear reactions and nuclear astrophysics with studies on nuclear 
level densities,
transition  strength densities, and parity/time-reversal violation 
(for example, see
\cite{applsSDT}).
The significance of the method is related to the fact that low-order energy
moments over a certain domain of single-particle states, such as the energy
centroid of an interaction (its average expectation value) and the deviation
from  that average, yield valuable information about the interaction that is of
fundamental importance
\cite{Potbhare77,HalemaneKD78,CounteeDHK81,French72,DraayerR83a,Ratcliff71_DBV79_SKK87}
without the need for carrying out large-dimensional matrix diagonalization and
with little to no limitations due to the dimensionality of the vector space.
Note that if one were to include higher-order energy moments, one 
would gradually
obtain more detailed results that, in principle, would eventually
reproduce those of a conventional microscopic calculations.

We compare three modern interactions, namely, the \CDB~\cite{MachleidtSS96M01},
\CDBt~\cite{PopescuSVN05} and \Gm~\cite{HonmaOBM04}, based on realistic
nucleon-nucleon potentials, as well as two model interactions with pairing and
quadrupole terms, which typically dominate in nuclei.
\CDB~ is a charge-dependent one-boson-exchange nucleon-nucleon ($NN$)
potential that is one of  the most accurate in
reproducing the available proton-proton and neutron-proton scattering
data. We employ the two-body
matrix elements of the effective interaction derived from \CDB~ for $0\hbar
\omega $ no-core shell model (NCSM) calculations in the $fp$ shell.
In addition, the
\CDBt~interaction introduces phenomenological isospin-dependent
central terms plus a tensor force with strengths  and
ranges determined in no-core $0\hbar\omega $ shell model calculations
to achieve an  improved description of the
$A=48$ Ca, Sc and Ti isobars. The \Gm~ effective interaction is
obtained from a realistic G-matrix interaction
based  on the Bonn-C potential \cite{Gint} by  adding empirical
corrections determined through systematic fitting to
experimental  energy data in the
$fp$ shell.

Several detailed reviews of  the nuclear shell model and its
applications have been published recently
\cite{CaurierMNPZ05,Brown01,OtsukaHMSU01} that delve into related key physics
issues that we explore.  However, the present study is novel and 
includes $fp$-shell
interactions, which have been developed since those reviews were completed.

\section{Theoretical Framework}
The theory of spectral distributions (or statistical spectroscopy) is
well documented in the literature
\cite{FrenchR71,ChangFT71,HechtDraayer74,French72,Kota79,ChangDW82}.
Spectral distribution theory combines important features, the most significant
of which are as follows:
\begin{enumerate}
\item The theory provides a precise measure, namely, the correlation
coefficient, for the
overall similarity of two interactions.  Literally the correlation
coefficient is a measure of the
extent to which two interactions ``look like" (are correlated with)
one another. In this respect,
correlation coefficients can be used to extract information how well
pairing/rotational features are
developed in realistic interactions, which may differ substantially
from an individual comparison of
pairing/quadrupole interaction strengths
\cite{SDV06}.

\item It gives an exact prescription for identifying the {\it pure} zero-
(centroid), one- and two-body parts of an interaction under a given
space partitioning. Therefore,
major properties follow:
\begin{enumerate}
      \item The correlation coefficients are independent of the
interaction centroids. (A direct
comparison of two-body matrix elements provided by $NN$ potentials
may be misleading,
especially when the averages of the interactions differ considerably.)

      \item  The pure one-body part of an interaction, the so-called induced
single-particle energies, is naturally identified in the framework of spectral
distribution theory and is indeed the average monopole interaction (compare to
\cite{HonmaOBM04}). As such it influences the evolution of the shell structure,
shell gaps and binding energies \cite{Otsuka01}.

      \item  The pure two-body part is essential for studies of detailed
property-defining two-body  interactions beyond strong mean-field effects.
\end{enumerate}

\item The correlation coefficient concept can be propagated
straightforwardly beyond the defining
two-nucleon system to derivative systems with larger numbers of nucleons
\cite{ChangFT71} and higher values of isospin \cite{HechtDraayer74}.
This, in addition to
the two-nucleon information provided by alternative approaches
(e.g.,\cite{DufourZ96}), yields
valuable overall information, without a need for carrying out
extensive shell-model calculations,
about the universal properties of a two-body interaction in shaping
many-particle nuclear systems.

\end{enumerate}

For a scalar  $\alpha =n$ (isospin-scalar $\alpha =n,T$) spectral 
distribution\footnote{For $n$ particles
distributed over $4\Omega$ single-particle states, {\it scalar} (or 
{\it isospin-scalar})
distribution is called the spectral
distribution averaged over the ensemble of all
$n$-particle states (of isospin $T$) associated with
the \Un{}{4\Omega} group
structure (or $\Un{}{2\Omega}
\otimes \Un{}{2}_T$).} the correlation coefficient between two Hamiltonian
operators $H$ and $H^\prime$ is defined as
\begin{equation}
\zeta ^\alpha _{H,H^\prime }=\frac{\langle (H^\dagger -\langle
H^\dagger \rangle
^\alpha ) (H^\prime-\langle H^\prime \rangle ^\alpha )
\rangle ^\alpha }{\sigma _H \sigma _{H^\prime}}
=\frac{\langle H^\dagger H^\prime\rangle ^\alpha -\langle H^\dagger \rangle
^\alpha \langle H^\prime \rangle ^\alpha }{\sigma _H \sigma _{H^\prime}},
\label{zeta}
\end{equation}
where the ``width" of $H$ is the positive square root of the
variance,
\begin{equation}
(\sigma ^\alpha _{H})^2=\langle (H-\langle H\rangle ^\alpha )^2
\rangle ^\alpha
=\langle H^2 \rangle ^\alpha -(\langle H \rangle ^\alpha )^2,
\label{sigma}
\end{equation}
and $\langle \cdots \rangle ^\alpha$ denotes an average value  related
to the trace of an operator divided by the dimensionality of the
space. The significance
of a positive correlation coefficient is given by Cohen \cite{Cohen88_03} and
later revised to the following table:
\begin{table} [h]
\caption{Interpretation of a correlation coefficient. \label{tab:cc}}
\smallskip
\begin{small}\centering
\begin{tabular*}{\textwidth}{@{\extracolsep{\fill}}crrrrrrrrr}
\hline  \noalign {\smallskip}
trivial  & small     & medium   & large    & very large & nearly perfect &
perfect \\
0.00-0.09 & 0.10-0.29 & 0.30-0.49 & 0.50-0.69 & 0.70-0.89   & 0.90-0.99      &
1.00\\
\hline
\end{tabular*}
\end{small}
\end{table}

   From a geometrical perspective, in spectral distribution theory
every interaction is associated with a vector and the correlation
coefficient $\zeta $ (Eq. \ref{zeta})
defines the angle (via a normalized scalar product) between two vectors of
length $\sigma $ (Eq. \ref{sigma}). Hence, $\zeta _{H,H^\prime}$
gives the normalized projection of $H$
onto the $H^\prime$ interaction (or $H^\prime$ onto $H$). In
addition, $(\zeta _{H,H^\prime})^2$
gives the percentage of $H$ that reflects the characteristic
properties of the $H^\prime$
interaction.

\section{Understanding the Nuclear Interaction in Many-nucleon Systems }

We compare the three modern interactions, $H_0=\{ $\CDB, \CDBt, \Gm 
$\}$, based on
realistic nucleon-nucleon potentials, and two model pairing and 
quadrupole isoscalar
interactions in the $fp$ region by means of the theory of spectral 
distributions. The
first model interaction, \Hsp, is the most general 
\Spn{4}-dynamically symmetric
interaction for a system of $n$ valence nucleons in a $4\Omega
$-dimensional space \cite{SGD03stg,SGD04} with two-body antisymmetric
$JT$-coupled matrix elements for $\{r\le (s,t);\ t\le u\}$ orbits
\cite{SDV06},
\begin{eqnarray}
W_{rstu}^{JT}=-G_0\frac{\sqrt{\Omega _r \Omega _t}}{\Omega }\delta
_{(JT),(01)}
\delta_{rs}\delta_{tu} -\{-E_0[(-)^T+\half]+C\}\delta_{rt}\delta_{su},
\label{W0me}
\end{eqnarray}
where $ G_0=G+\frac{F}{3},\ E_0=(\frac{E}{2\Omega}+\frac{D}{3})$,
$G,F,E,D$ and $C$ are interaction strength parameters (see Table I in 
Ref.\cite{SGD04}
for parameter estimates). The \spn{4} algebraic structure is exactly 
the one needed in
nuclear
\IASs~ to describe
proton-proton ($pp$), neutron-neutron $(nn)$ and proton-neutron
($pn$) isovector pairing
correlations [the first term in (\ref{W0me})], while the second term includes
isoscalar $pn$ force related to the $E_0$ isospin symmetry term.

In addition, we construct an extended pairing plus quadrupole model, $H_M$, by
including an additional traceless quadrupole-quadrupole interaction
that being symmetric under \SU{}{3} breaks the \Spn{4} symmetry,
\begin{equation}
H_M=H_{\spn{4}}+\HQ,\
H_Q= -\frac{\chi }{2} Q \cdot Q ,
\label{HM}
\end{equation}
where $\chi $ is the only parameter in the present analysis and is 
determined by
optimum correlation coefficients. The \HQ-term is the part of the pure
two-body $H_Q(2)$ interaction that is not contained in (is orthogonal to) the
\Spn{4} interaction\footnote{Such a Hamiltonian (Eq.
\ref{HM}) does not affect the centroid of \Hsp~ because
$\HQ$ is traceless and hence
preserves the
shell structure that is built into \Hsp~ and established by a
harmonic oscillator
potential and as a result is favored in many studies
\cite{HalemaneKD78,CounteeDHK81,DraayerR83b}.}.
This is because the \Spn{4} interaction
contains a nonnegligible part of the quadrupole-quadrupole interaction
that is revealed by the correlation between $H_Q$ and \Hsp.
Namely, in the scalar
case  it is $15\%$ ($\flevel$), $29\%$ ($\fFive $)
and $29\%$ ($\plevels$), and for the T=1 part of the interactions in
the isospin-scalar case, it is
$34\%$ ($\flevel$), $58\%$ ($\fFive$) and $58\%$ ($\plevels$). This is
probably  one of the reasons why the
\Spn{4} model interaction turns out to work rather well despite no
explicit appearance of the quadrupole-quadrupole interaction.

The present study focuses on the weaker but property-defining two-body part of
the interactions, such as $H_0(2)$, in the \flevel \cite{SDV06} and \upfp-shell
\cite{SDV06b} domains. Such a partitioning of the $fp$ oscillator shell follows
naturally from a splitting of these two regions by a strong 
spin-orbit interaction.
Several correlation coefficients are of particular interest. Specifically, the
overall similarity between a realistic interaction and the extended
pairing$+$quadrupole model interaction in the (isospin-) scalar case 
is estimated by the
$\zeta_{H_0(2),H_M}^{n(T)}$ correlation, while the capability of a realistic
interaction to describe rotational collective motion, and hence
to reproduce rotational bands and enhanced electric quadrupole
transitions, can be detected
via its correlation to the full $H_Q(2)$ quadrupole-quadrupole
two-body interaction, $\zeta_{H_0(2),H_Q(2)}^{n(T)}$. The isospin-scalar
space partitioning is where the ability of a realistic
interaction to form correlated pairs and hence reproduce prominent pairing
gaps is detected via $\zeta_{H_0(2),\Hsp}^{n,T}$. In the present analysis, the
$\zeta_{H_0(2),H_Q(2)}^{n(T)}$ and $\zeta_{H_0(2),\Hsp}^{n,T}$ 
correlation coefficients
are independent of the quadrupole/pairing interaction strength parameters.

\begin{figure}[th]
\centerline{\epsfig{file=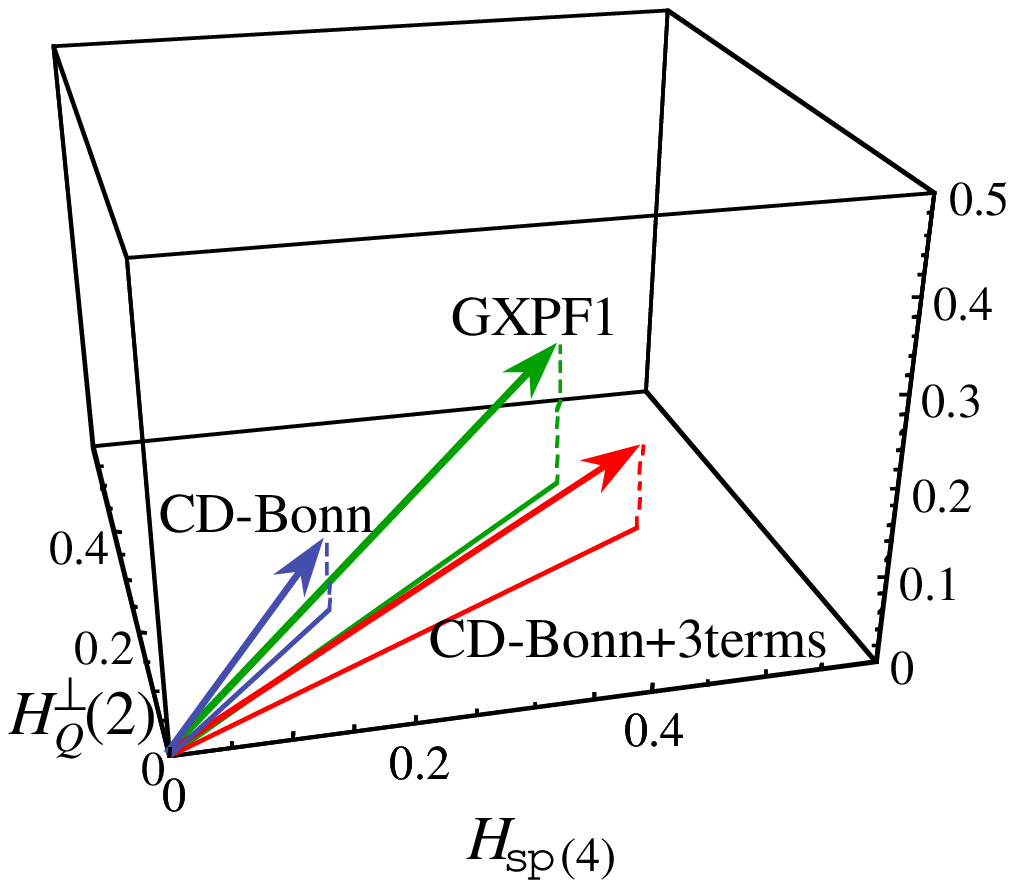,width=44mm}}
\caption{ Geometrical representation of the $T=1$
\CDB~ (light blue),
\CDBt~(red) and \Gm~(green) interactions, in an abstract operator 
space, where the
horizontal
plane is spanned by the orthogonal linear operators, the pure
two-body \Hsp~ and
$\HQ$ model Hamiltonians, both linearly independent of the residual interaction
operators represented by the vertical axis. The orientation of the
vectors remains the same
for any particle number $n \ge 2$ and for all $T=n/2$ cases.}
\label{Vs}
\end{figure}

In the detailed case of isospin-scalar distribution in the \flevel orbit, the
\CDB, \CDBt~ and \Gm~ interactions are found to  contain
on average $59\%$, $77\%$, and $78\%$, respectively, of the
pairing$+$quadrupole interaction. This percentage  goes up to $91\%$, 
$97\%$, and
$92\%$, respectively, for the highest possible isospin
group of states for all the nuclei  with valence protons and neutrons
occupying the \flevel
shell (Fig. \ref{Vs}). For these states, the strongest correlation was
observed between the \CDBt~ and the
pairing$+$quadrupole model interaction, where other types of
interactions accounted for in the
realistic  interaction represent only $3\%$ of it. They constitute $8\%$
of the \Gm~ interaction, and $9\%$ of \CDB. While both
interactions, \CDBt~ and \Gm, exhibit a well-developed pairing character
compared to \CDB, the latter
appears to build up more (less) rotational collective features that
are outside of the
scope of the $T=1$ ($T=0$) \Spn{4} interaction.

\begin{figure}[th]
\centerline{\epsfig{file=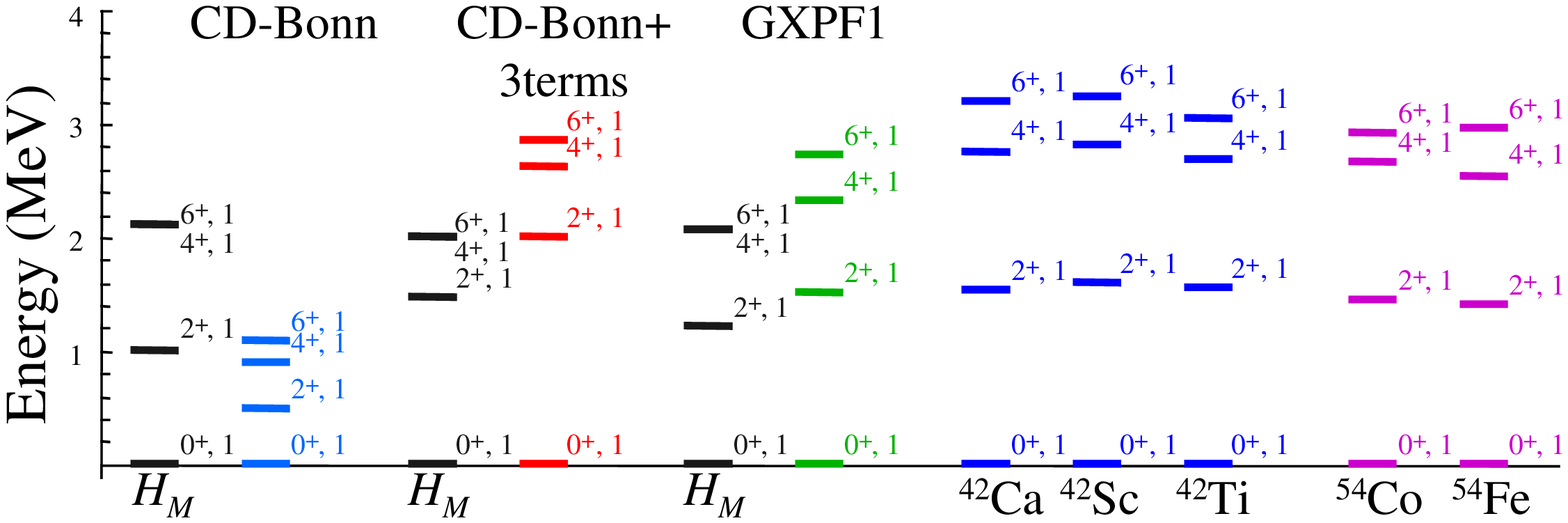,width=80mm}}
\caption{ Energy spectra of $T=1$ states predicted by
the \CDB~ (light
blue), \CDBt~
(red) and \Gm~ (green) interactions. Each is compared to the model
Hamiltonian $H_M$ (black) with $\chi =0.071,\ 0.036$ and $0.055$,
respectively. For comparison, the
experimental $T=1$ energy
spectra of the $A=42$  Ca, Sc, Ti isobars (blue) and $A=54$ Co and Fe
isobars (magenta)
are  also shown.}
\label{TscalarSpectra}
\end{figure}

The abovementioned strong correlation coefficients,
$\zeta_{H_0,H_M}^{n,T=n/2}$, imply that both realistic and $H_M$ interactions
are expected to yield  energy spectra of a similar pattern. Indeed, for these
cases the pairing$+$quadrupole model interaction appears to be a very good
approximation that provides a reasonable description of the energy spectra of
the nuclei in the
\flevel level (Fig. \ref{TscalarSpectra}).

\begin{figure}[th]
\centerline{\epsfig{file=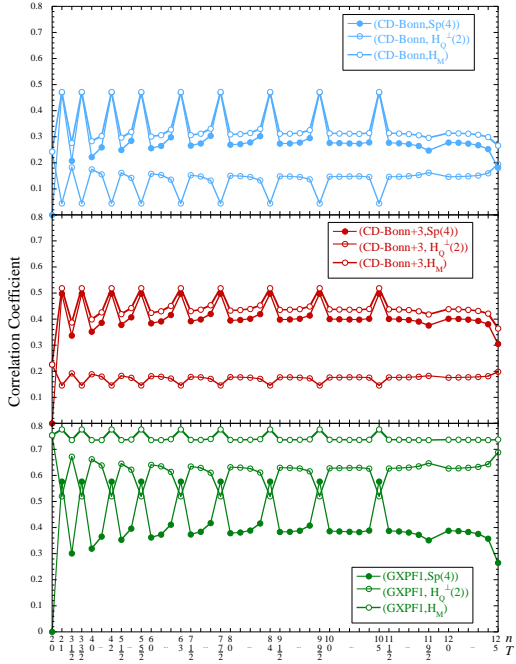,width=68mm}}
\caption{Correlation coefficients of the pure two-body
\CDB~(blue), \CDBt~(red) and \Gm~(green) interactions with \Hsp~ (filled
symbols), \HQ~(transparent symbols) and $H_M$ (empty symbols) in the
\upfp~ shell for the isospin-scalar distribution. For each valence
particle number, $n$, the isospin $T$ varies as
$\frac{n}{2},\frac{n}{2}-1,\dots, 0(\frac{1}{2})$. The figures are symmetric
with respect to the sign of $n-2\Omega$.\label{upfp}}
\end{figure}

In the \upfp~ domain the outcome yields strong correlation of the pure
two-body \Gm~ interaction with the pairing+quadrupole extended
model in both scalar and isospin-scalar distributions (Fig. \ref{upfp}).
Particularly, the outcomes show very good scalar-distribution
correlations of \Gm~ with the \Spn{4}
dynamically symmetric interaction and with the $H_Q(2)$ quadrupole-quadrupole
interaction (Table \ref{tab:upfpS}).
Rotational features within many-nucleon
systems in the \upfp~ domain are found to be more fully developed for
\Gm~ and less for \CDBt~ and \CDB. The isospin-scalar space
partitioning demonstrates a tendency in
\Gm~ towards the formation of correlated pairs in the highest
possible isospin groups of states (Fig. \ref{upfp}).
\begin{table} [th]
\caption{Correlation coefficients for many-nucleon systems of the $H_0(2)$
pure two-body part of the \CDB, \CDBt~and \Gm~ interactions  with 
\Hsp~ and \HQ, with
the pure two-body full quadrupole-quadrupole interaction, $H_Q(2)$, 
and with the
extended pairing+quadrupole model interaction \HM~ (Eq. \ref{HM}).
\label{tab:upfpS}}
\smallskip
\begin{small}\centering
\begin{tabular*}{\textwidth}{@{\extracolsep{\fill}}crrrrrrrrr}
\hline  \noalign {\smallskip}
& \CDB    & \CDBt                               &   \Gm\\
\hline \hline
$\zeta _{H_0(2),\Hsp}$    & 0.55    &  0.50    &  0.65  \\
$\zeta _{H_0(2),\HQ}$     & 0.14    &  0.20    &  0.51  \\
$\zeta _{H_0(2),H_Q(2)}$  & 0.28    &  0.33    &  0.67  \\
$\zeta _{H_0(2),H_M}$     & 0.57    &  0.54    &  0.83  \\
\hline
\end{tabular*}
\end{small}
\end{table}

The different extent to which the \Gm~ interaction compared to the
\CDB~and \CDBt~ interactions
reflects development of pairing correlations and collective
rotational modes in the \upfp~ domain
may be the reason why their pure two-body part do not correlate as
strongly as, for example, \CDB~and
\CDBt~ do. Namely, in the scalar case the pure two-body correlations
are $0.90$ (between \CDB~ and
\CDBt) and only $0.56$  (\CDB~ and \Gm) and $0.53$ (\CDBt~ and \Gm). In
the isospin-scalar case, the
correlations vary slightly with the particle number and isospin and
they are on average, $0.88$
(between \CDB~ and
\CDBt), $0.40$ (\CDB~ and \Gm), and $0.37$ (\CDBt~ and \Gm). In addition,
one can compare the
significant monopole influence of the three interactions, which is
very similar for all when
averaged over the isospin values. However, in the isospin-scalar
distribution, the correlation
coefficients involving the induced effective one-body contribution
differ between \Gm~ and the two \CDB~ interactions. Their behavior,
especially below mid-shell, reflects the similarity of the corresponding $T=0$
induced  single-particle energies and
the opposite signs of the corresponding $\lambda _{3/2}^{T=1}$ (for
$2p_{3/2}$) and  $\lambda _{5/2}^{T=1}$ (for
\fFive) pure one-body interactions.

Individual orbit analysis, including the  $\flevel$, $\fFive$,
$2p_{1/2}$, and $2p_{3/2}$ levels, shows considerably stronger correlation of
all the interactions with the pairing+quadrupole model interaction (up to
$0.8-1.00$) as well as in nuclear systems with more than two nucleons.
However, more prominent differences among the interactions appear in the
multi-$j$ \upfp~ domain especially concerning both \CDB~interactions. This may
indicate that the inter-orbit interactions do not respect strongly the
symmetries imposed in the model interactions. In addition, the interaction
strengths may differ from one orbit to another. While they do not affect
correlation coefficients in the singe-$j$ cases, their relative strength is of
a great importance for multi-$j$ analysis. In addition, the 
difference in the behavior
of \CDBt~ within both regions, \flevel~ and \upfp, may reflect the 
fact that this
interaction was determined through a reproduction of the energy 
spectrum and binding
energy of $A=48$ \flevel nuclei.

\begin{figure}[th]
\centerline{\epsfig{file=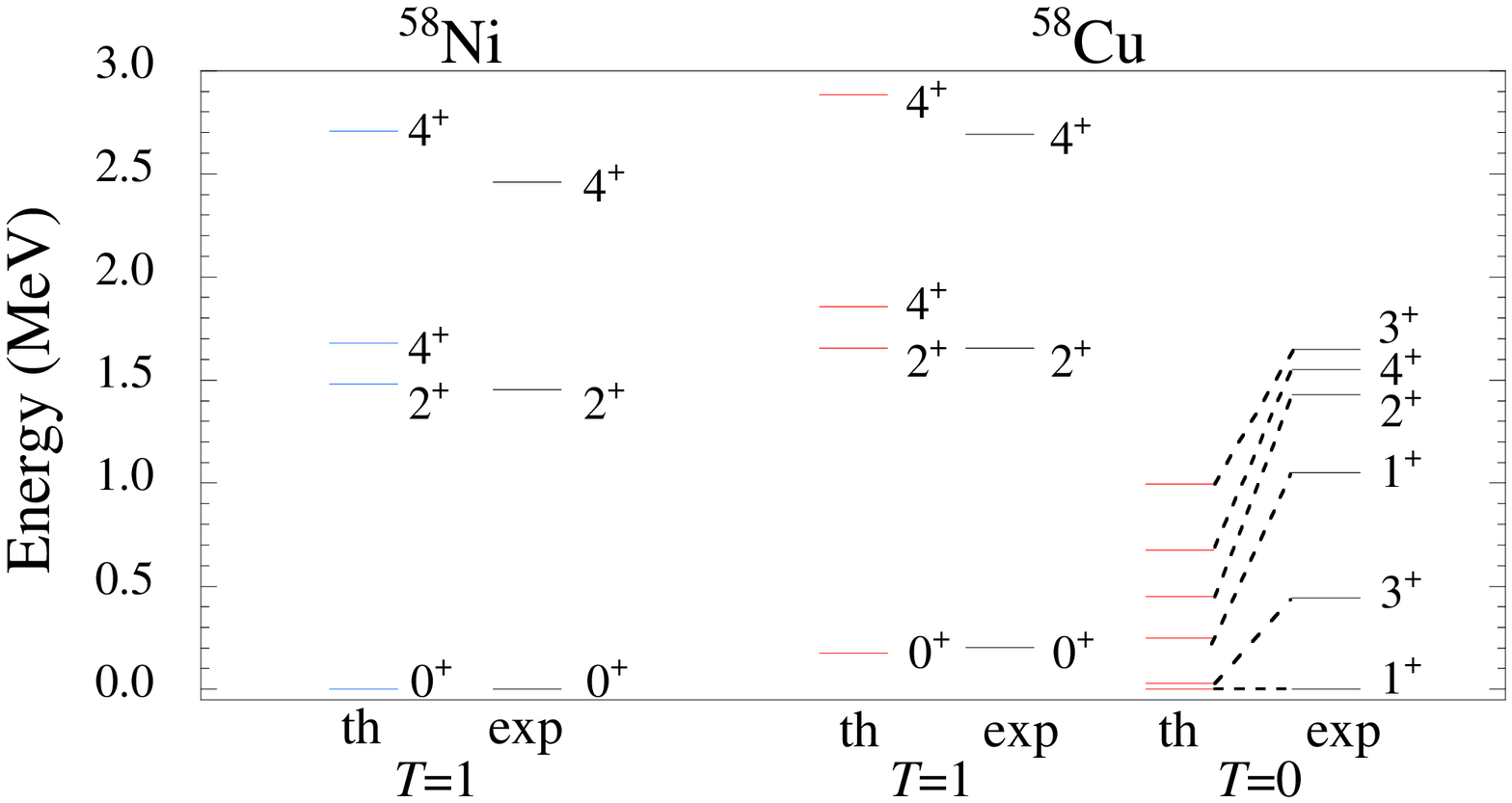,width=70mm}}
\caption{Theoretical (`th') low-lying energy spectra for $^{58}$Ni
(left, blue) and $^{58}$Cu (right,
red) compared to experiment (`exp', black). The theoretical
calculations are performed in the \fpg
major shell with the $H_M$ model interaction with
$\chi=0.027$ and with
single-particle energies derived from $^{57}$Ni experimental energy levels.
\label{enSpectra}}
\end{figure}

In the \upfp~ region, the extended $H_M$ pairing+quadrupole
interaction is strongly correlated with
the pure two-body \Gm~ interaction especially in the scalar
distribution (Table \ref{tab:upfpS}) and for this
reason can  be used as a good approximation. This is reflected in the
quite good agreement between
the experimental low-lying energy spectra of $^{58}$Ni and $^{58}$Cu
and the theoretical prediction
based on the $H_M$ model interaction with $\chi=0.027$ (see Eq. 
\ref{HM}) in the \fpg~
major shell (Fig.
\ref{enSpectra}).
\newline

In summary, the present study reveals important information, within  the
framework of spectral
distribution theory, about the types of forces that dominate  the $fp$-shell
\CDB, \CDBt~ and \Gm~ interactions in nuclei and their
ability to account for many-particle effects such
as the formation of correlated nucleon  pairs and enhanced quadrupole
collective modes. The results also illustrate that
interactions, which were found to strongly correlate, produced energy spectra
of a similar pattern.

\section*{Acknowledgments}
JPV would like to thank Sorina Popescu, Sabin Stoica and
Geanina Negoita for valuable discussions.
This work was supported by the US National Science Foundation, Grant
Numbers 0140300 \& 0500291, and the Southeastern Universities Research
Association, as well as partly performed under the auspices
of the US Department of
Energy by the University of California, Lawrence Livermore
National Laboratory under contract No.  W-7405-Eng-48
and under the auspices of grants DE-FG02-87ER40371 \&
DE--AC02--76SF00515.


\begin{thebibliography}{0}

\bibitem{FrenchR71} J. B. French and K. F. Ratcliff, {\it Phys. Rev.}
{\bf C3}, 94
(1971).

\bibitem{ChangFT71} F. S. Chang, J. B. French, and T. H. Thio, {\it Ann.
Phys.} (N.Y.)
{\bf 66}, 137 (1971).

\bibitem{DraayerOP75}  J. P. Draayer, J. B. French, V. Potbhare, and
S. S. M. Wong, {\it Phys. Lett.} {\bf 55B}, 263, 349 (1975); B. D. 
Chang and J. P.
Draayer, {\it Phys. Rev.} {\bf C20}, 2387 (1979).

\bibitem{Potbhare77} V. Potbhare, {\it Nucl. Phys.} {\bf A289}, 373
(1977).

\bibitem{HechtDraayer74} K. T. Hecht and J. P. Draayer, {\it Nucl.
Phys.} {\bf A223},
285 (1974).

\bibitem{SDV06}  K. D. Sviratcheva, J. P. Draayer, and J. P. Vary,
{\it Phys. Rev.} {\bf C73}, 034324 (2006).

\bibitem{SDV06b}  K. D. Sviratcheva, J. P. Draayer, and J. P. Vary,
submitted (2006).

\bibitem{SGD03stg}  K. D. Sviratcheva, A. I. Georgieva, and J. P. Draayer,
Phys. Rev. {\bf C69} 024313 (2004).

\bibitem{SGD04}  K. D. Sviratcheva, A. I. Georgieva, and J. P. Draayer,
Phys. Rev. {\bf C70} 064302 (2004).

\bibitem{Elliott} J. P. Elliott, {\it Proc. Roy. Soc.} (London) {\bf
A245}, 128 (1958); {\bf A245}, 562 (1958);
J. P. Elliott and M. Harvey, {\it Proc. Roy. Soc.} (London) {\bf
A272}, 557 (1963).

\bibitem{Draayer73}  J. P. Draayer, {\it Nucl. Phys.} {\bf A216}, 457 (1973).

\bibitem{HalemaneKD78} T. R. Halemane, K. Kar, and J. P. Draayer, {\it
Nucl. Phys.}
{\bf A311}, 301 (1978).

\bibitem{KotaPP80}  V. K. B. Kota, S. P. Pandya, and V. Potbhare, {\it
Nucl. Phys.} {\bf A349}, 397 (1980).

\bibitem{CounteeDHK81} C. R. Countee, J. P. Draayer, T. R. Halemane, and
K. Kar, {\it
Nucl. Phys.} {\bf A356}, 1 (1981).

\bibitem{applsSDT} J. B. French, V. K. B. Kota, A.  Pandey, and S.
Tomsovic, Ann. Phys. (N.Y.) {\bf 181}, 235 (1988); V. K. B. Kota and 
D. Majumdar, Z.
Phys. A {\bf 351}, 365 (1995);  Z. Phys. A {\bf 351}, 377 (1995); S. 
Tomsovic, M. B.
Johnson, A. Hayes, and J. D. Bowman, {\it Phys. Rev.} {\bf C62}, 
054607 (2000); J. M. G.
Gomez, K. Kar, V. K. B. Kota, R. A. Molina, and J. Retamosa,
{\it Phys. Lett.} {\bf B 567}, 251 (2003); M. Horoi, M. Ghita, and V. 
Zelevinsky, {\it
Phys. Rev.} {\bf C69}, 041307(R) (2004); N. D. Chavda, V. Potbhare, and V. K.
B. Kota, {\it Phys. Lett.}, {\bf A 326}, 47 (2004); V. K. B. Kota, 
{\it Phys. Rev.} {\bf
C71}, 041304(R) (2005); Y. M. Zhao, A. Arima, N. Yoshida, K. Ogawa, N.
Yoshinaga, and V. K. B. Kota, {\it Phys. Rev.} {\bf C72}, 064314 (2005).

\bibitem{French72}  J. B. French, in {\it Dynamic Structure of
Nuclear States}, ed. D. J. Rowe {\it et al.} (Univ. of Toronto Press,
Toronto, 1972), p.154.

\bibitem{DraayerR83a}  J. P. Draayer and G. Rosensteel  {\it
Phys. Lett.} {\bf 124B}, 281 (1983); G. Rosensteel and J. P. Draayer,
{\it Nucl.
Phys.} {\bf A436}, 445 (1985).

\bibitem{Ratcliff71_DBV79_SKK87}  K. F. Ratcliff, {\it Phys. Rev.} {\bf
C3}, 117 (1971); B. J. Dalton, W. J. Baldridge, and J. P. Vary,
{\it Phys. Rev.} {\bf C20}, 1908 (1979); S. Sarkar, K. Kar, and V. K. 
B. Kota, {\it
Phys. Rev.} {\bf C36}, 2700 (1987).

\bibitem{MachleidtSS96M01} R. Machleidt, F. Sammarruca, and Y. Song,
{\it Phys. Rev.}
{\bf C53}, R1483 (1996); R. Machleidt, {\it Phys. Rev.} {\bf C63},
024001 (2001).

\bibitem{PopescuSVN05} S. Popescu, S. Stoica, J. P. Vary, and P.
Navratil, to be published.

\bibitem{HonmaOBM04} M. Honma, T. Otsuka, B. A. Brown, and T. Mizusaki, {\it
Phys. Rev.} {\bf C69}, 034335 (2004).

\bibitem{Gint} M. Hjorth-Jensen, T. T. S. Kuo, and E. Osnes, {\it Phys. Rep.}
{\bf 261}, 125 (1995).

\bibitem{CaurierMNPZ05} E. Caurier, G. Martinez-Pinedo, F. Nowacki,
A. Poves, and A. P. Zuker, {\it
Rev. Mod. Phys.} {\bf 77}, 427 (2005).

\bibitem{Brown01} B. A. Brown, {\it Prog. Part. Nucl. Phys.} {\bf
47}, 517 (2001).

\bibitem{OtsukaHMSU01} T. Otsuka, M. Honma, T. Mizusaki, N. Shimizu,
and Y. Utsuno,
{\it Prog. Part. Nucl. Phys.} {\bf 47}, 319 (2001).

\bibitem{Kota79} V. K. B. Kota, {\it Phys. Rev.} {\bf C20}, 347
(1979); Fortran Programs for Statistical Spectroscopy Calculations.

\bibitem{ChangDW82}  B. D. Chang, J. P. Draayer, and S. S. M. Wong,  {\it
Comput. Phys. Commun.} {\bf 28}, 41 (1982).

\bibitem{Otsuka01} T. Otsuka \etal, {\it Phys. Rev. Lett.} {\bf 87},
082502 (2001).

\bibitem{DufourZ96} M. Dufour and A. P. Zuker, {\it Phys. Rev.} {\bf
C54}, 1641 (1996).

\bibitem{Cohen88_03} J. Cohen, {\it Statistical Power Analysis for the
Behavioral Sciences} (Lawrence Erlbaum Associates, Hillsdale, New Jersey,
1988); J. Cohen, P. Cohen, S. G. West, and L. S. Aiken, {\it Applied
multiple regression/correlation analysis for the behavioral sciences}, 2nd ed.
(Lawrence Erlbaum Associates, Hillsdale, New Jersey, 2003).

\bibitem{DraayerR83b}  J. P. Draayer and G. Rosensteel,  {\it
Phys. Lett.} {\bf 125B}, 237 (1983).

\end{thebibliography}
\end{document}